\documentclass[prd,superscriptaddress,amsfonts,amssymb,amsmath,showpacs,twocolumn]{revtex4-2}
\usepackage{bm}
\usepackage{amsfonts}
\usepackage{latexsym}
\usepackage[latin1]{inputenc}
\usepackage{graphicx}
\usepackage{amsmath}
\usepackage{palatino}
\usepackage{mathpazo}
\usepackage{textcomp}
\usepackage{float}
\usepackage{booktabs}
\usepackage{dcolumn}
\usepackage{hyperref}

\hypersetup{colorlinks,citecolor=red}

\usepackage{amsmath}
\usepackage{xcolor}
\usepackage{orcidlink}
\usepackage[caption=false]{subfig}
\usepackage{commath}
\begin{document}

\color{black}

\title{Cosmological observational constraints on the power law $f(Q)$ type modified gravity theory}

\author{Sanjay Mandal\orcidlink{0000-0003-2570-2335}}
\email{sanjaymandal960@gmail.com}
\affiliation{Faculty of Mathematics and Computer Science, Transilvania University, Iuliu Maniu Str. 50, 500091 Brasov, Romania}

\author{Sneha Pradhan\orcidlink{0000-0002-3223-4085}}
\email{snehapradhan2211@gmail.com}
\affiliation{Department of Mathematics, Birla Institute of Technology and
Science-Pilani, Hyderabad Campus, Hyderabad-500078, India}

\author{P.K. Sahoo\orcidlink{0000-0003-2130-8832}}
\email{pksahoo@hyderabad.bits-pilani.ac.in}
\affiliation{Department of Mathematics, Birla Institute of Technology and
Science-Pilani, Hyderabad Campus, Hyderabad-500078, India}

\author{Tiberiu Harko\orcidlink{}}
\email{tiberiu.harko@aira.astro.ro}
\affiliation{Department of Theoretical Physics, National Institute of Physics and Nuclear Engineering (IFIN-HH),077125 Bucharest, Romania}
\affiliation{Department of Physics, Babes-Bolyai University, 1 Mihail Kogalniceanu Street, 400084 Cluj-Napoca, Romania}

\date{\today}

\begin{abstract}

In modern cosmology, the curiosity of ultimately understanding the nature of the dark energy controlling the recent acceleration of the Universe motivates us to explore its properties by using some novel approaches. In this work, to explore the properties of dark energy we adopt the modified $f(Q)$ gravity theory, where the non-metricity scalar $Q$, emerging from Weyl geometry,  plays the dynamical role. For the function $f(Q)$ we adopt the functional form $f(Q)=Q+ 6\gamma\,H_0^2(Q/Q_0)^n$, where $n,\, \gamma,\, H_0$ and $Q_0$ are constants.  Then, we test our constructed model against the various observational datasets, such as the Hubble, and the Pantheon+SHOES samples, and their combined sample, through the Markov Chain Monte Carlo (MCMC) statistical analysis. We also employ the parameter estimation technique to constrain the free parameters of the model. In addition, we use the constrained values of the model parameters to explore a few implications of the cosmological model. A detailed comparison of the predictions of our model with the $\Lambda$CDM model is also performed. In particular, we discuss in detail some cosmographic parameters, like the deceleration, the jerk, and the snap parameters, as well as the behavior of the dark energy and matter energy densities to see the evolution of various energy/matter profiles. The $Om$ diagnostics is also presented to test the dark energy nature of our model, as compared to the standard $\Lambda$CDM paradigm. Our findings show that the considered version of the non-metric $f(Q)$ type modified gravity theory, despite some differences with respect to the $\Lambda$CDM paradigm, can still explain the current observational results on the cosmological parameters, and provide a convincing and consistent account for the accelerating expansion of the Universe.


\end{abstract}

\maketitle

\tableofcontents

\section{Introduction}\label{sec:1}

In present day cosmology, one of the primary objective is to explain the accelerating expansion of our Universe, an effect whose existence was extensively proven, and investigated, over the past two decades \cite{SCP,AG}. To understand the accelerating phase of the Universe, one must either modify Einstein's General Relativity, or add a new exotic component, called dark energy (DE) to the universe's energy budget. DE is an exotic fluid type component,  having a negative pressure that causes gravity to behave in a repulsive manner at large cosmological scales \cite{LM}. The equation-of-state parameter $\omega(z)$, defined as the ratio of the fluid's pressure to its energy density, is usually employed to characterize the dynamical features of DE. The most straightforward hypothesis to explain the cosmological observations is to assume that dark energy is a cosmological constant, with the parameter of the equation of state given by the redshift independent $\omega=-1$. The cosmological constant, together with the assumption of the existence on the Universe of a called dark matter component are the conceptual basis of the $\Lambda$CDM cosmological paradigm.  Alternative cosmological models that depart from the conventional $\Lambda$CDM model, but still predict an accelerating expanding Universe include braneworld models \cite{RM}, K-essence, quintessence, and non-minimally coupled scalar fields \cite{BR,CA,AV,LAU, Ha1}, modified gravity \cite{SC,SN,SA,WH,AAs,Ha2,Ha3,Ha4,Ha5,Ha6,Ha7}, anisotropic universes \cite{OA,VM,WV}, interacting dark energy \cite{LA,TC,JL}, and many others \cite{RC,VS,IZ,SM1,VP,GG,JLC,Ade}.\\

Based on the equivalence principle, the view of the gravitational force as a manifestation of the curvature of the space-time became the dominant paradigm for the understanding of the gravitational force. This assumption implies that the gravitational interaction, and the geometry of the space-time, are completely determined by the nature of the matter fields. The Ricci scalar curvature plays a vital role in the curved space-time geometry. The Ricci scalar curvature $R$ is the basic quantity from which the standard Einstein's general relativity has been built initially in a Riemannian geometry, where the torsion and the non-metricity do vanish. Although it is well known that Einstein's general relativity provides an outstanding description of the local gravitational phenomena, at the level of the Solar System, the theory has been theoretically challenged by specific observational evidence coming from the realization that the Universe is accelerating,  and from the galactic phenomenology that is usually explained by postulating the existence  of dark matter. These observations suggest that for explaining the gravitational dynamics and galactic and extra-galactic scales one should go beyond the standard formalism of general relativity.

The simplest way to construct extensions of general relativity is to include either an additional component in the Einstein-Hilbert Lagrangian, or to modify the structure of the Einstein-Hilbert gravitational Lagrangian (the Ricci scalar) itself.  These approaches have led to many important extensions of general relativity, including  $f(R)$ gravity \cite{AA}, $f(G)$ gravity \cite{ADe}, $f(P)$ gravity\cite{CEE}, Horndeski scalar-tensor theories\cite{HST} etc. However, from a general differential geometric perspective, by taking into account the affine properties of a manifold, the curvature is not the only geometric object that may be used within a geometrical framework to construct gravitational theories. Torsion and nonmetricity are two other essential geometric objects connected to a metric space, along with the curvature. They can be used to obtain the $f(T)$ and the $f(Q)$ gravity theories, respectively.

In the current paper, we are going to describe the current accelerated expansion of the Universe, and the observational data, through a specific modified gravity theory, the symmetric teleparallel gravitation theory, alternatively called $f(Q)$ gravity. The $f(Q)$ gravity was first proposed by Nester and Yo \cite{Nester}, and later extended by Jimenez et al. \cite{JBJ}. In $f(Q)$ gravity the non-metricity $ Q$, originating from the Weyl geometric background,  describes the gravitational interaction in a flat geometry, in which the curvature vanishes. $f(Q)$ gravity was extensively used to investigate the cosmological evolution of the Universe. By considering the $f(Q)$ Lagrangian of the theory as polynomial function in the redshift $z$, Lazkoz et al.\cite{RL} obtained an important number of restrictions on $f(Q)$ gravity.  This investigation demonstrated that viable $f(Q)$ models have coefficients comparable to those of the GR model, specifically the $\Lambda$CDM model. In the work \cite{BBN}, researchers proposed a new model in which they showed their model immediately passes BBN restrictions since it does not show early dark energy features, and the change of the effective Newton's constant lies within the bounds of observation. Another new cosmological model has been studied by the same research group \cite{BBN2} related to BBN formalism in order to extract the constraints on various classes of $f(Q)$ models. To investigate if this new formalism offers any workable alternatives to explain the Universe's late-time acceleration, the validity of various models at the background level was investigated. Several observational probes for the analysis have been employed, including the expansion rates of the early-type galaxies, Type Ia supernovae, Quasars, Gamma Ray Bursts, Baryon Acoustic Oscillations, and Cosmic Microwave Background distance priors. It turns out that the novel approach proposed in $f(Q)$ gravity offers a different perspective on constructing modified, observationally reliable cosmological  models.

The exploration of stellar models in the $f(Q)$ modified gravity theory has been performed in \cite{MKO}, in which observational restrictions in the context of $f(Q)$ gravity are obtained from the study of compact general relativistic objects. Focusing on a particular model in $f (Q)$ gravity, Frusciante \cite{FR} found that while it is identical to the $\Lambda$CDM model at the background level, it exhibits novel and measurably different signatures at the level of linear perturbations. By examining the external and internal solutions for compact stars, Lin and Zhai \cite{ZH} investigated the application of $f(Q)$ gravity to the static spherically symmetric configurations and illustrated the consequences of the $f(Q)$ gravity theory. Mandal et al.\cite{SM} explored the dark energy parameters for the non-linear and power-law $f(Q)$ models that depict the observable behavior of the cosmos. Jimenez et al.\cite{JIM} investigated the modified gravity theories based on nonlinear extensions of the nonmetricity scalar, and they examined several interesting baseline cosmologies (including accelerating solutions related to inflation and dark energy), and assessed how cosmic perturbations behaved. Harko et al.\cite{HAR} considered an extension of $f(Q)$ gravity, by considered the effects of a non-minimal coupling between geometry and matter. Several cosmological applications of the theory were considered, by obtaining the generalized Friedmann equations (the cosmological evolution equations),  and by imposing specific functional forms of the function $f(Q)$, such as power-law and exponential dependence of the nonminimal couplings. A full theory in which nonmetricity couples to matter, called $f(Q,T)$ gravity, where $T$ is the trace of the matter energy-momentum tensor,  was introduced and developed in \cite{Xu1} and \cite{Xu2}. Some astrophysical implications of the $f(Q,T)$ theory were investigated in \cite{Yang}. The inclusion of the torsion in the formalism of theories with geometry-matter coupling was considered in \cite{HAR1}.  In addition, for studying various types of energy restrictions for the investigation of the logarithmic and polynomial functions in the $f(Q)$ gravity, Mandal et al.\cite{MAN} used cosmographic quantities to reconstruct the proper structure of the $f(Q)$ function. The evolution of matter perturbations in the modified $f (Q)$ gravity was investigated by Khyllep et al. \cite{KH}, who also considered the power-law structure of the cosmic perturbations.

It is the goal of the present paper to consider a detailed investigation, in the framework of $f(Q)$ gravity, of a specific cosmological model, obtained by assuming a simple power law form of the $f(Q)$ function, $f(Q)=Q+ \gamma 6H_0^2(Q/Q_0)^n$, where $n,\gamma $ and $Q_0=6 H_0^2$ are constants. After writing down the generalized Friedmann equations, an effective dark energy model can be constructed. As for the parameter of the equation of state of the dark energy we assume a specific, redshift dependent form. In order to test the predictions of the model we have adopted several numerical techniques, including the MCMC fitting,  which allow us to study the observational implications of this modified $f(Q)$ gravity model, which gives us the possibility of constraining the cosmological model parameters, using various observational datasets.

This manuscript is organized in the following manner. We start with the presentation of the basic formulation of the $f(Q)$ gravity in Section \ref{sec:2}. We present the basic assumptions and ideas of a specific $f(Q)$ type cosmological model in Section~\ref{sec:3}. Thereafter, in Section~\ref{sec:4}, we present the different observational samples, the numerical methods, and we present the data analysis outputs. Moreover, we discuss the obtained results in detail. In addition, in Section \ref{sec:5}, we explore the behavior in our model of various cosmological quantities, like the deceleration parameter, jerk and snap parameters, and the dark energy and dark matter densities, respectively. Finally, we discuss and conclude our results in Section \ref{sec:6}.

\section{Brief review of the $f(Q)$ gravity theory}\label{sec:2}

The basic idea of the $f(Q)$ theory is that gravitational phenomena can be fully described in the Weyl geometry \cite{Nester}, in which the metric conditions is not anymore satisfied, and the covariant divergence of the metric tensor is given by
\begin{equation}
\nabla _\lambda g_{\mu \nu}=Q_{\lambda \mu \nu},
\end{equation}
where  $Q_{\lambda \mu \nu}$ is called the nonmetricity. The scalar non-metricity, given by
\begin{equation}
    Q\equiv -g^{\mu\nu}\left(L^\alpha _{\ \beta\nu}L^\beta_{\ \nu\alpha}-L^{\alpha}_{\ \beta\alpha}L^{\beta}_{\ \mu\nu}\right),
\end{equation}
plays a fundamental role in the theory,
where $L^\lambda_{\ \mu\nu}$ is defined as,
\begin{equation}
    L^{\lambda}_{\ \mu\nu}=-\frac{1}{2}g^{\lambda\gamma}\left(Q_{\mu\gamma\nu}+Q_{\nu\gamma\mu}-Q_{\gamma\mu\nu}\right).
\end{equation}

Now, we introduce the action for the $f(Q)$ gravity theory, given by \cite{JBJ},
\begin{eqnarray}\label{x1}
S=\int\left[\frac{1}{2}f(Q)+\mathcal{L}_m\right]\sqrt{-g}d^{4}x,
\end{eqnarray}
where $f(Q)$ is a general function of the non-metricity scalar $Q$, $g$ represents the determinant of the
metric $g_{\mu \nu}$, and $\mathcal{L}_m$ is the matter
Lagrangian density. The non-metricity tensor is given as,
\begin{eqnarray}
Q_{\alpha \mu \nu}=\nabla_{\alpha}g_{\mu \nu}=-L^{\rho}_{\alpha \mu}g_{\rho \nu}-L^{\rho}_{\alpha \nu}g_{\rho \mu}.
\end{eqnarray}
The following two equations give the expressions of the non-metricity tensor's two independent traces
 \begin{eqnarray}
 Q_{\alpha}=Q^{~\beta}_{\alpha ~\beta},~~ \tilde{Q}_{\alpha}=Q^{\beta}_{~~\alpha \beta},
 \end{eqnarray}
 while the deformation term is given by
\begin{eqnarray}
L^{\alpha}_{\mu \nu}=\frac{1}{2}Q^{\alpha}_{\mu \nu}-Q^{~~~\alpha}_{(\mu \nu)}.
\end{eqnarray}
Moreover, the nonmetricity scalar $Q $ is obtained as,
 \begin{eqnarray}
Q=-g^{\mu\nu}(L^{\alpha}_{\beta \nu}L^{\beta}_{\mu \alpha}-L^{\beta}_{\alpha \beta}L^{\alpha}_{\mu \nu})=-P^{\alpha \beta \gamma}Q_{\alpha \beta \gamma}.
 \end{eqnarray}
Here, $P^{\alpha \beta \gamma}$ is the non-metricity conjugate, and is defined as
\begin{eqnarray}
P^{\alpha}_{~~\mu\nu}=\frac{1}{4}\left[-Q^{\alpha}_{~~\mu\nu}+2Q^{\alpha}_{(\mu\nu)}-Q^{\alpha}g_{\mu \nu}-\tilde{Q}^{\alpha}g_{\mu \nu}-\delta^{\alpha}_{(\mu}Q_{\nu})\right].\nonumber\\
\end{eqnarray}

The field equation of the $f(Q)$ gravity theory is obtained by varying (\ref{x1}) with respect to $g_{\mu\nu}$, and it takes the following form:
\begin{eqnarray}
-\frac{2}{\sqrt{-g}}\nabla_a (\sqrt{-g}f_QP_{\mu \nu}^{\alpha})+\frac{1}{2}g_{\mu\nu}f\\ +f_Q(P_{\nu}^{\alpha\beta}Q_{\mu \alpha \beta}-2P^{\alpha \beta}_{~~\mu}Q_{\alpha\beta\nu})=\kappa T_{\mu \nu},
\end{eqnarray}
where $f_Q=\frac{\partial f}{\partial Q}$, and
the energy-momentum tensor $T_{\mu \nu}$ is given by
\begin{eqnarray}\label{tmu1}
T_{\mu \nu}&=&-\frac{2}{\sqrt{-g}}\frac{\delta \sqrt{-g}\mathcal{L}_m}{\delta \sqrt{g_{\mu \nu}}},
\end{eqnarray}

By varying the action with respect to the affine connection, the following equation can be obtained:
\begin{eqnarray}
\nabla_{\mu}\nabla_{\nu}(\sqrt{-g}f_QP^{\mu \nu}_{~~~\alpha})=0.
\end{eqnarray}

Within the framework of $f(Q)$ gravity, the field equations guarantee the conservation of the energy-momentum tensor, and given the choice of $f(Q)=Q$, the Einstein equations are retrieved.

\section{The cosmological model}\label{sec:3}

The standard Friedmann-Lemaitre-Robertson-Walker line element, which describes our flat, homogeneous, and isotropic Universe, is given by,
\begin{equation}
    ds^2 = - dt^2 + a^2(t) (dx^2 + dy^2 +dz^2).
\end{equation}
Here $t$ is the cosmic time, and $x,\, y,\, z$ denote the Cartesian co-ordinates. Moreover, $a(t)$ is the cosmic scale factor. The Hubble parameter $H(t)$ is defined by $H(t)=\frac{\dot a}{a}$, where $\dot{a}$ denotes the derivative of $a$ with respect to the cosmic time $t$. Moreover, we introduce the cosmological redshift $z$ defined as $1+z=1/a$.

\subsection{The generalized Friedmann equations}

For the FLRW geometry we get the non-metricity scalar as $Q=6 H^2$. We consider the matter content of the Universe as consisting of a perfect and isotropic fluid, with energy-momentum tensor given by
\begin{eqnarray}
    T_{\mu \nu} &=& (p+\rho) u_{\mu} u_{\nu}+ p g_{\mu \nu},
\end{eqnarray}
where $p$ and $\rho$ are the pressure and the energy density of the fluid, $u_{\mu}$ is the four velocity vector normalized according to $u^{\mu} u_{\mu}=-1$.

Now we are considering the splitting of $f(Q)$ as $f(Q) = Q + F(Q)$. By considering the FLRW metric, we get two Friedmann equations as \cite{J1,J2}
\begin{equation}
    3H^2 = \rho+\frac{F}{2}-Q F_Q,\label{eq:3}
\end{equation}
\begin{equation}
    (2QF_{QQ}+F_{Q}+1)\dot{H}+ \frac{1}{4}(Q+2QF_Q-F) = -2p \label{eq:4}
\end{equation}
where $F_{Q}=\frac{dF}{dQ}$ and $F_{QQ}=\frac{d^2{F}}{dQ^2}$.

In the above equation (\ref{eq:3}), the energy density ($\rho$) can be written as $\rho=\rho_m+\rho_r$ where $\rho_m,\, \rho_r $ \, are the energy density for dark matter and radiation, respectively. Similarly, we can write $p=p_r+p_m$. The standard matter distribution satisfies the conservation equation given by,
\begin{equation}\label{eq:c1}
    \frac{d\rho}{dt} + 3 H (1+\omega) \rho=0.
\end{equation}

In  Eq.~(\ref{eq:c1}), the equation of state parameter (EoS) for matter, $\omega$, takes different values for different matter sources, like baryonic matter, and radiation. As for the expression of $Q$, and its time derivative,  they are related to the Hubble parameter by the important relations
\begin{equation}
Q=6H^2, \dot{Q}=12H \dot{H} .
\end{equation}

\subsection{The equation of state of the dark energy}

On the  other hand,  to describe the features of dark energy, due to the lack of  precision of the current data, and our lack of theoretical understanding of dark energy, extracting the value of EoS of dark energy from observational data is particularly difficult. Under these circumstances, one must parameterize $\omega_{de}$ empirically, usually using two or more free parameters, to probe the dynamical evolution of dark energy. The Chevallier-Polarski-Linder (CPL) model \cite{XD} is the most popular and thoroughly studied among all the parametrization forms of dark energy EoS. The simplest form of the CPL model can be written as,
\begin{equation}\label{wde}
    \omega_{de}(z) = \omega_0 + \omega_a \frac{z}{1+z}.
\end{equation}

In the above expression, $z$ is the redshift, $\omega_0$ denotes the present-day value of EoS $\omega(z)$, and $\omega_a$ characterizes its dynamics. The main reason for considering such a parametrization form is to resolve the divergence property of the linear form $\omega (z) = \omega_0 + \omega_a z$ at high redshifts.

In addition, the CPL parametrization has a number of advantages, as mentioned by Linder \cite{EV}, including a manageable two-dimensional phase space, well-behaved and bounded behavior for high redshifts, high accuracy in reconstructing numerous scalar field equations of state, a straightforward physical interpretation, etc.

Though it has the above mentioned benefits, there are some drawbacks to the CPL model. The CPL model only properly describes the past expansion
history, but cannot describe the future evolution, since $\abs{\omega_{de}(z)}$ increases and finally diverges
as $z$ approaches $-1$. The EoS is bound between $\omega_0+\omega_a$ and $\omega_0$ from the infinite past to the present.

\subsection{The generalized Friedmann equations in the redshift space}

In general, for isotropic and homogeneous spatially flat FLRW cosmologies in the presence of radiation, non-relativistic matter, and an exotic fluid with an equation of state $p_{de}=\omega_{de} \, \rho_{de} $, the Friedmann equations \eqref{eq:3}, \eqref{eq:4} becomes
\begin{equation}\label{eq:8}
    3H^2=\rho_r +\rho_m+\rho_{de},
\end{equation}
\begin{equation}\label{eq:8a}
    2\dot{H}+ 3H^2= -\frac{\rho_r}{3}-p_m-p_{de},
\end{equation}
where $\rho _r$, $\rho_m$, and $p_m$ are the energy densities of the radiation and matter components, $p_m$ is the matter pressure, while $\rho_{de}$ and $p_{de}$ are the DE's density and pressure contribution due to the geometry, given by
\begin{equation}\label{rde}
    \rho_{de}= \frac{F}{2}-Q\, F_Q,
\end{equation}
\begin{equation}
    p_{de}= 2\dot{H} (2Q F_{QQ}+F_Q)-\rho_{de}.
\end{equation}

In the following we assume that the matter pressure, be it baryonic, or dark matter, can be neglected. From Eqs.~(\ref{eq:8}) and (\ref{eq:8a}) we obtain immediately the global conservation equation
\begin{equation}
\frac{d}{dt}\left(\rho _r+\rho_m+\rho_{de}\right)+3H\left(\frac{4\rho_r}{3}+\rho_m+\rho_{de}+p_{de}\right)=0.
\end{equation}

 When there are no interactions between the three fluids, the energy densities satisfy the following differential equations
\begin{eqnarray}
\dot{\rho}_r+4H\rho_r&=&0,\label{eq:e1}\\
\dot{\rho}_m+3H \rho_m&=&0,\label{eq:e2}\\
\dot{\rho}_{de}+3H(1+\omega_{de})\rho_{de}&=&0 \label{eq:e3}.
\end{eqnarray}

The dark energy equation of state $\omega_{de}$ can be written as the function of $F(Q)$ and its derivatives as

\begin{equation}
  \omega_{de}=\frac{p_{de}}{\rho_{de}}  =-1+\frac{4\dot{H}(2QF_{QQ}+F_{Q})}{F-2QF_{Q}}.
\end{equation}
 From Eqs.~(\ref{eq:e1}) and (\ref{eq:e2}), one can quickly get the evolution of the pressureless matter and of radiation, namely, $\rho_m \propto \frac{1} {a(t)^{3}} $
and $\rho_r \propto \frac{1} {a(t)^{4}}$.

Moreover, by using the relationship between redshift ($z$) and the universe scale factor $a(t)$ \big[$a(t)=\frac{1}{1+z}$\big], we can represent the relationship between the redshift and the cosmic time as,
\begin{equation}
    \frac{d}{dt}=\frac{dz}{dt}\frac{d}{dz}=-(1+z) H(z) \frac{d}{dz}.
\end{equation}

Now, for the present cosmological study of the $f(Q)$ gravity,  we are considering one particular form of $F(Q)$, with
\begin{equation}\label{fq}
F(Q)=6\gamma\,H_0^2\left(\frac{Q}{Q_0}\right)^n,
\end{equation}
where $H_0$, $\gamma$, $n$ and $Q_0$ are constants. The motivation for choosing this form is that the Friedmann equations represent a system of ordinary differential equations, and we can find power-law and exponential types of solutions for these types of equations. Therefore, we have considered the power-law form in our study. With the adopted functional form of $f(Q)$ we obtain first
\begin{eqnarray}\label{rde}
\rho_{de}&=&\frac{F}{2}-Q\,F_{Q}=\frac{\alpha Q^n}{2}-Qn\alpha Q^{n-1}=\alpha  \left(\frac{1}{2}-n\right)Q^n\nonumber\\
&=& 6\gamma\,H_0^2\left(\frac{1}{2}-n\right)\left(\frac{Q}{Q_0}\right)^n=6\gamma\,H_0^2\left(\frac{1}{2}-n\right)\left(\frac{H}{H_0}\right)^{2n},\nonumber\\
\end{eqnarray}
where we have denoted $\alpha=6\gamma\,H^2_{0}/Q_{0}^n$, and $Q_0=6H_0^2$. Then for the derivative of the dark energy we obtain the expression
\begin{eqnarray}
\dot{\rho}_{de}&=&n\,\alpha\, Q^{n-1}\left(\frac{1}{2}-n\right)\dot{Q}\nonumber\\
&=&12n\gamma\,H_0^2\left(\frac{1}{2}-n\right)\left(\frac{H}{H_0}\right)^{2n}\frac{\dot{H}}{H}.
\end{eqnarray}

We substitute now the expressions of the dark energy, and of its derivative, into the conservation equation (\ref{eq:e3}), together with the CPL parametrization of the parameter of the dark energy equation of state. Hence, by also taking into account the relation between $H$ and $Q$, we obtain
\begin{equation}
2n\,\frac{\dot{H}}{ H}+3H\left(1+\omega_{de}\right)=0,
\end{equation}
leading, in the redshift space, to the first order differential equation
\begin{equation}
-2 n\,(1+z) \frac{dH}{dz}+3H\left(1+\omega_0+\omega_a \frac{z}{1+z}\right)=0,
\end{equation}
or
\begin{equation}
-n(1+z)\frac{d}{dz}H^2+3\left(1+\omega_0+\omega_a \frac{z}{1+z}\right)H^2=0,
\end{equation}
with the general solution given by
\begin{equation}\label{hde}
    H^2(z)=C_1^2 (1+z)^{\frac{3(1+\omega_o+\omega_a)}{n}} e^{\frac{3 \omega_a }{n(1+z)}},
\end{equation}
where $C_1$ is an arbitrary constant of integration, which we determine so that $H^2(0)=H_0^2$, giving $C_1^2=H_0^2e^{-3\omega _a/n}$. Hence we obtain
\begin{equation}
H^2(z)=H_0^2 (1+z)^{\frac{3(1+\omega_o+\omega_a)}{n}}  e^{-\frac{3\omega_a z}{n(1+z)}}.
\end{equation}

Now using  \eqref{hde} in \eqref{rde}, we obtain for the dark energy density $\rho_{de}$ the expression
\begin{equation}
    \rho_{de}(z)= 3\gamma\, (1-2n) H_0^{2}(1+z)^{3(1+\omega_o+\omega_a)} e^{\frac{-3 \omega_a z}{(1+z)}}.
\end{equation}

Alternatively, we can obtain the same result by using the considered equation of state, which gives first
\begin{eqnarray}
  \omega_{de}&=&\frac{P_{de}}{\rho_{de}}  =\frac{2\dot{H}(2QF_{QQ}+F_{Q})-\rho_{de}}{\frac{F}{2}-QF_{Q}}\nonumber\\
  &=&-1+\frac{4\dot{H}(2QF_{QQ}+F_{Q})}{F-2QF_{Q}}.
\end{eqnarray}

With the help of the CPL parametrization we successively obtain
\begin{equation}
-1- \frac{4 \dot{H} n}{Q}=\omega_0+\omega_a \frac{z}{1+z},
\end{equation}
and
\begin{equation}
-\frac{2}{3}n(1+z)\frac{dH}{dz}\frac{1}{H}=-\left[1+\omega_0+\omega_a \frac{z}{1+z}\right],
\end{equation}
respectively, with the solution of the above differential equation given again by Eq.~(\ref{hde}).

Additionally, the matter density ($\rho_m$) and radiation density ($\rho_r$) can be written in terms of redshift function $z$ as,
\begin{gather}
    \rho_m \propto (1+z)^3 \,\,; \quad \rho_r \propto (1+z)^4
\end{gather}

Consequently, the Friedmann equation (\ref{eq:8}) reads,
\begin{widetext}
    \begin{equation}
        \begin{gathered}\label{eq:h1}
            3H^2(z)=\rho_{r0}(1+z)^4+\rho_{m0}(1+z)^3+3 \gamma\, (1-2n) H_0^{2}(1+z)^{3(1+\omega_o+\omega_a)} e^{\frac{-3 \omega_a z}{(1+z)}}, \\
            \frac{H^2(z)}{ H_0^2}=\Omega_{r0}(1+z)^4+\Omega_{m0}(1+z)^3+\gamma\, (1-2n) (1+z)^{3(1+\omega_o+\omega_a)} e^{\frac{-3 \omega_a z}{(1+z)}}.
        \end{gathered}
\end{equation}
\end{widetext}

In the equation (\ref{eq:h1}), the suffix $0$ represents the present day value of the corresponding quantity. $H_0$ is the current Hubble value (at $z=0$) of our present Universe.

Finally, we are going to introduce the energy density parameters, defined as
\begin{equation}
    \begin{gathered}
        \Omega_{m}=\frac{\rho_m}{3 H^2},\,\,\, \Omega_{r}=\frac{\rho_r}{3H^2},\,\,\,\Omega_{de}=\frac{\rho_{de}}{3H^2}
    \end{gathered}
\end{equation}

\section{Observational Data}\label{sec:4}

In this Section we discuss the methodology, and the various observational samples used to constrain the parameters $H_0,\, \Omega_{m0},\,\omega_0,\, \omega_a,\, n, \gamma$ of the considered cosmological model. In particular, we use a Markov Chain Monte Carlo (MCMC) method to do the statistical analysis, and to obtain the posterior distributions of the parameters. The data analysis part is done by using the \textit{emcee} package in Python. The best fits of
the parameters are maximized by using the probability function
\begin{eqnarray}
    \mathcal{L}\propto \exp(-\chi^2/2),
\end{eqnarray}
 where $\chi^2$ is the pseudo chi-squared function \cite{baye}. More details about the $\chi^2$ function for various date samples are discussed in the following subsections.

\subsection{Cosmic Chronometer (CC) Sample}

For the Cosmic Chronometer (CC) Sample, we used 31 points of Hubble samples, collected from the differential age (DA) approach in the redshift range $0.07 < z < 2.42$. The complete list of this sample is collectively presented in \cite{mand}. The chi-square function for the Hubble sample is defined as
\begin{equation}
\chi_{CC}^2=\sum_{i=1}^{31}\frac{[H_i^{th}(\theta_s,z_i)-H_i^{obs}(z_i)]^2}{\sigma_{CC}^2(z_i)}
\end{equation}
where $H_i^{obs}$ denotes the observed value, $H_i^{th}$ denotes the Hubble's theoretical value, $\sigma_{z_i}$ denotes the standard error in the observed value, and $\theta_s= (H_0, \Omega_{m0},\,\omega_0, \omega_a, n, \gamma)$ is the cosmological background parameter space. In addition, we use the following \textit{priors} to our analysis, which we present in Table~\ref{table1}.
\begin{table}[!htbp]
    \centering
    \caption{Priors for the parameter space $H_0, \Omega_{m0},\,\omega_0, \omega_a, n,\gamma$.}
    \begin{tabular}{|c| c|}
    \hline\hline
    Parameter & prior\\\hline
        $H_0$ & (60,80) \\
        $\Omega_{m0}$ & (0,1)\\
        $\omega_0$ & (-2,2)\\
        $\omega_a$ & (-2,2)\\
        $n$ & (-1,1)\\
        $\gamma$ & (-1,1)\\
        \hline
    \end{tabular}

    \label{table1}
\end{table}

\begin{figure*}[!htbp]
    \centering
    \includegraphics[scale=.6]{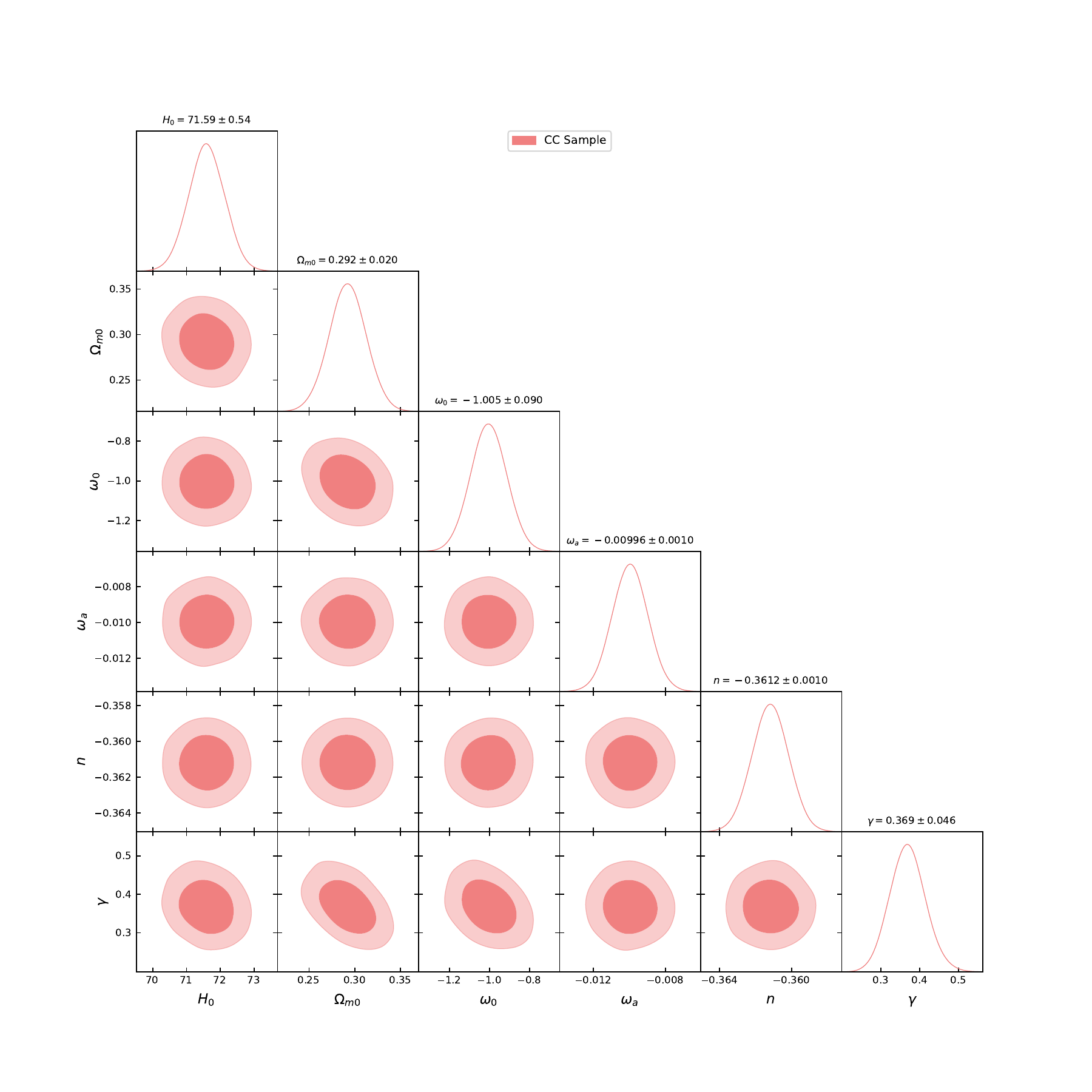}

    \caption{The marginalized constraints on the parameters $H_0, \Omega_{m0},\,\omega_0, \omega_a, n, \gamma$ of our model using the Hubble sample. The dark orange shaded regions presents the $1-\sigma$ confidence level (CL), and the light orange shaded regions present the $2-\sigma$ confidence level. The constraint values for the parameters are presented at the $1-\sigma$ CL.}
    \label{fig1}
\end{figure*}

\begin{figure*}[!htbp]
    \centering
    \includegraphics[scale=.6]{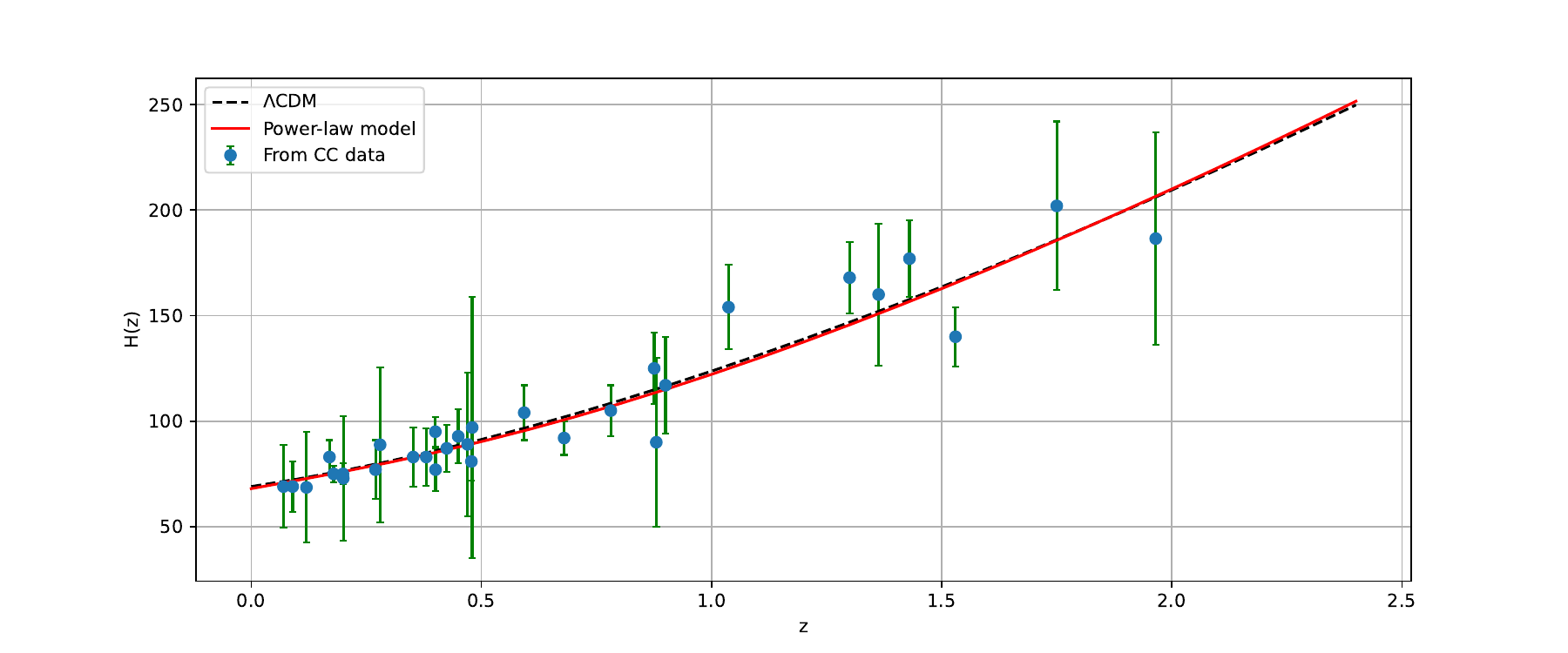}

    \caption{The red line represents the Hubble parameter profile of the power-law model $f(Q)$ model with the constraint values of $H_0, \Omega_{m0},\,\omega_0, \omega_a, n, \gamma$. The blue dots with the green bars represent the CC sample, and the black dotted line represents the Hubble parameter profile of the $\Lambda$CDM model. }
    \label{fig2}
\end{figure*}

  In our MCMC analysis, we used $100$ walkers and $1000$ steps to find out the fitting results. The $1-\sigma$ and $2-\sigma$ CL contour plot are presented in Fig.~\ref{fig1}, and the numerical results are presented, for the CC sample, in Table~\ref{table2}. With the mean constrain value of the free parameters, we present the Hubble parameter profile for the CC sample, together with the $\Lambda$CDM behavior, in Fig.~\ref{fig2}.

\subsection{Type Ia Supernovae Sample}

 Supernovae samples are a powerful indicator for exploring the background geometry and properties of the Universe. In this analysis, we adopt the largest SNe Ia sample published to date, the Pantheon+SHOES sample, which consists of 1701 light curves of 1550 spectroscopically confirmed SNe Ia across 18 different surveys \cite{pan+}. The Pantheon+SHOES sample significantly increases the number of observations relative to the Pantheon data at low redshifts, and covers the redshift range $z \in  [0.00122, 2.26137]$. It is the successor of Pantheon sample \cite{pantheon}. The chi-square function is defined as,
\begin{equation}\label{4b}
\chi^2_{SN}=\sum_{i,j=1}^{1701}\bigtriangledown\mu_{i}\left(C^{-1}_{SN}\right)_{ij}\bigtriangledown\mu_{j}.
\end{equation}
Here $C_{SN}$ is the covariance matrix \cite{pan+}, and
\begin{align*}\label{4c}
\quad \bigtriangledown\mu_{i}=\mu^{th}(z_i,\theta)-\mu_i^{obs}.
\end{align*}
is the difference between the observed value of distance modulus, extracted from the cosmic observations, and its theoretical values, calculated from the model, with the given parameter space $\theta$. $\mu_i^{th}$ and $\mu_i^{obs}$ are the theoretical and observed distance modulus, respectively.

The theoretical distance modulus $\mu_i^{th}$ is defined as
\begin{equation}
\mu_i^{th}(z)=m-M=5\log D_l(z)+25,
\end{equation}
 where $m$ and $M$ are apparent and the absolute magnitudes of a standard candle, respectively. The luminosity distance $D_l(z)$ is defined as
 \begin{equation}
D_l(z)=(1+z)\int_{0}^z\frac{dz^\ast}{H(z^\ast)}.
\end{equation}

\begin{figure*}[!htbp]
    \centering
    \includegraphics[scale=.6]{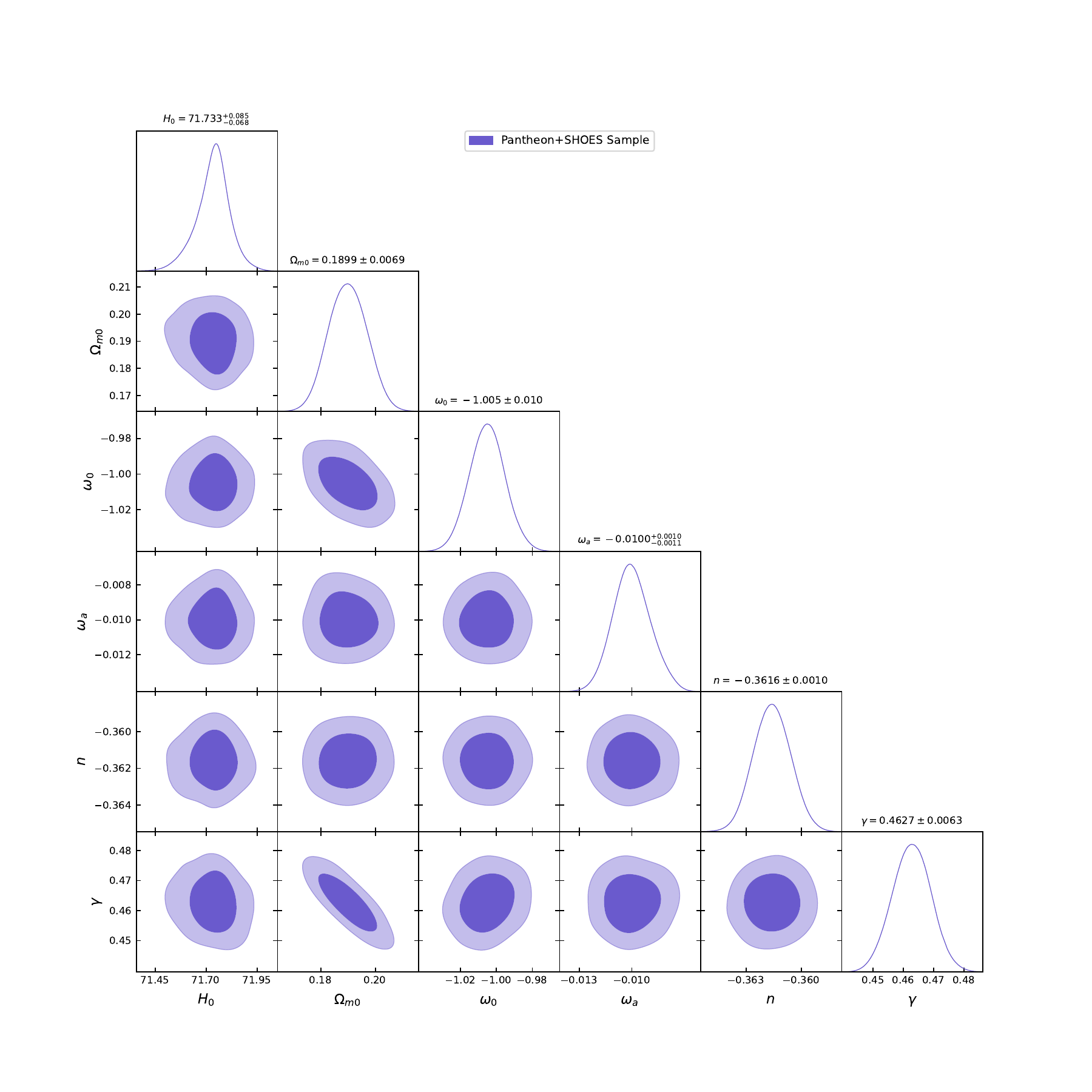}

    \caption{The marginalized constraints on the parameters $H_0, \Omega_{m0},\,\omega_0, \omega_a, n, \gamma$ of our model using Pantheon+Shoes sample. The dark blue shaded regions present the $1-\sigma$ confidence level (CL), and light blue shaded regions present the $2-\sigma$ confidence level. The constraint values for the parameters are presented at the $1-\sigma$ CL.}
    \label{fig3}
\end{figure*}

\begin{figure*}[!htbp]
    \centering
    \includegraphics[scale=.6]{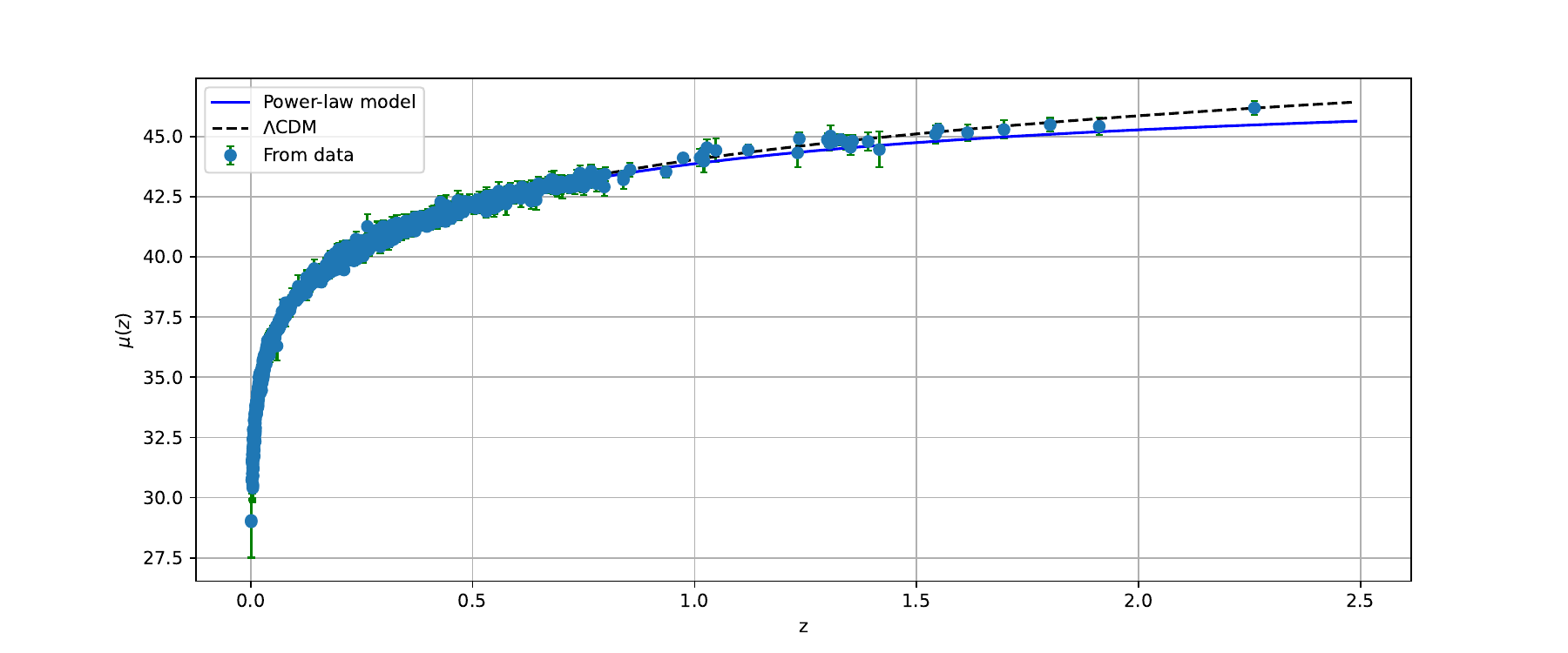}

    \caption{The blue line represents the distance modulus profile of the power-law $f(Q)$ model with the constraint values of $H_0, \Omega_{m0},\,\omega_0, \omega_a, n,  \gamma$. The blue dots with the green bars represent the Pantheon+SHOES sample, and the black dotted line represents the distance modulus profile of the $\Lambda$CDM model.}
    \label{fig4}
\end{figure*}

To run the MCMC code, we used the same \textit{priors}, number of walkers, and steps, which have been used in the CC sample. The $1-\sigma$ and $2-\sigma$ CL contour plot is presented in Fig.~\ref{fig3}, and the numerical results for the Pantheon+Shoes sample are presented in Table~\ref{table2}. With the mean constraint value of the free parameters, we present the distance modulus parameter profile with the Pantheon+SHOES sample and the $\Lambda$CDM model in Fig.~\ref{fig4}.

\subsection{CC + Type Ia Supernovae Sample}

\begin{figure*}[!htbp]
    \centering
    \includegraphics[scale=0.6]{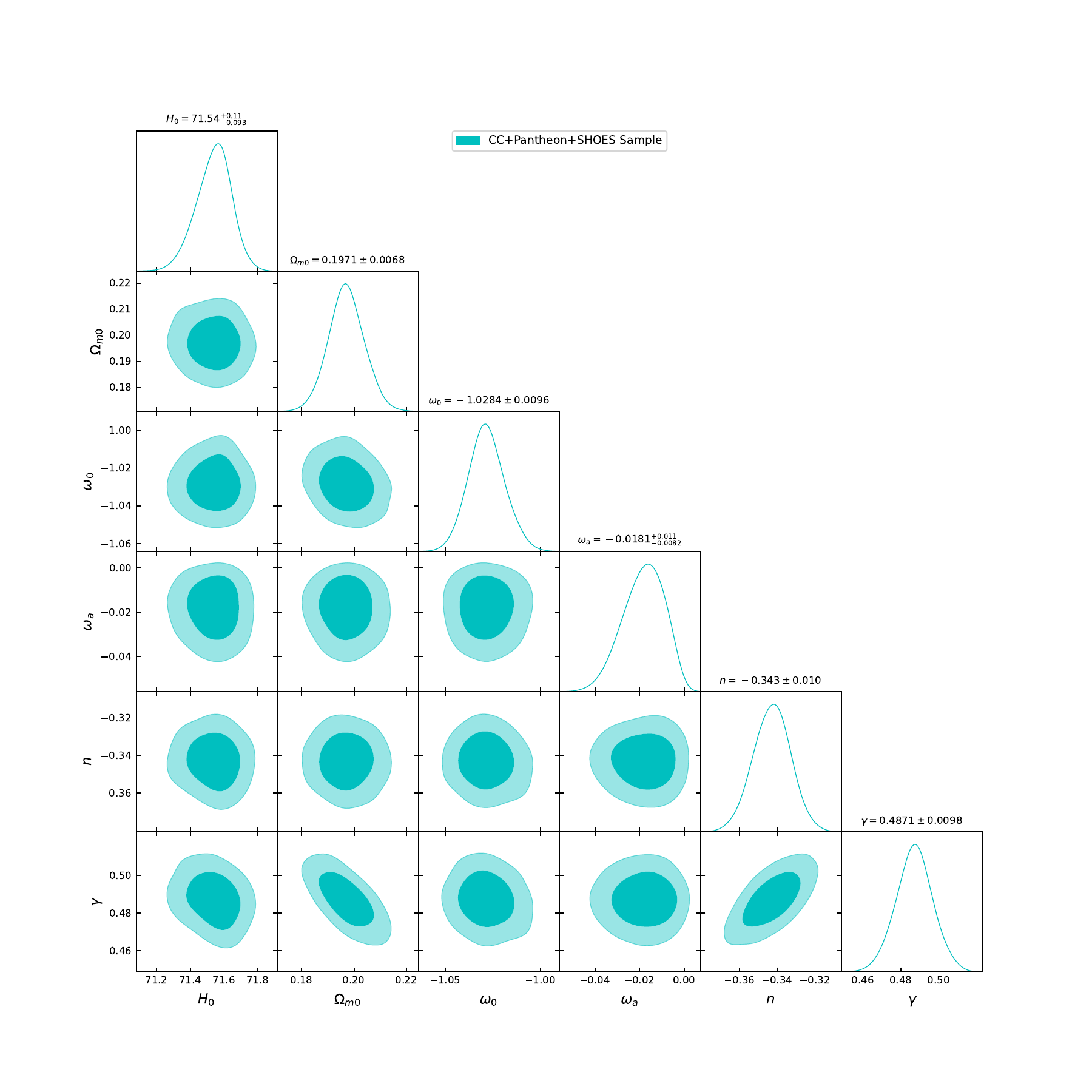}

    \caption{The marginalized constraints on the parameters $H_0, \Omega_{m0},\,\omega_0, \omega_a, n, \gamma$ of our model using the Hubble+Pantheon sample. The dark-shaded regions present the $1-\sigma$ confidence level (CL), and the light-shaded regions present the $2-\sigma$ confidence level. The constraint values for the parameters are presented at the $1-\sigma$ CL.}
    \label{fig5}
\end{figure*}

To perform both the CC and Type Ia supernovae samples together, we use the following Chi-square function
\begin{equation}
    \chi^2_{CC+SN}=\chi^2_{CC}+\chi^2_{SN}.
\end{equation}
The marginalized constraints on the parameters included in the parameter space $\theta$ are presented in Fig.~\ref{fig5}. The numerical results are presented in Table \ref{table2}.

\begin{center}
\begin{table*}[!htbp]
\caption{Marginalized constrained data of the parameters $H_0$,\,\,$\Omega_{m0}$, $\omega_0,\,\,\, \omega_a,\, \gamma$ and $n$ for different data samples with 68\% and 95\% confidence level. }
		    \label{table2}
		\begin{ruledtabular}
		    \centering
		    \begin{tabular}{|c c c c c c c|}
				 Model & $H_0$ &$\Omega_{m0}$ & $\omega_0$& $\omega_a$ & $n$&$\gamma$\\
    	\hline
     & & & \textbf{$68\%$ limits} & & &\\
     & & & CC sample & &&\\
    $\Lambda$CDM &$68.80\pm 0.94 $ & $0.318\pm 0.034  $&-&-&-&-\\
     Power-law &  $71.59\pm 0.54 $ & $0.292\pm 0.020$ & $-1.005\pm 0.090 $ & $-0.00996\pm 0.0010 $ & $-0.3612\pm 0.0010 $& $0.369\pm 0.046 $\\

     &&&&&&\\
     & & & Pantheon+SHOES sample & &&\\
      $\Lambda$CDM &$72.33\pm 0.28             $ & $0.383\pm 0.022            $&-&-&-&-\\
Power-law &  $71.733^{+0.085}_{-0.068}  $& $0.1899\pm 0.0069          $ & $-1.005\pm 0.010           $& $-0.0100^{+0.0010}_{-0.0011}$ & $-0.3616\pm 0.0010         $& $0.4627\pm 0.0063          $\\

&&&&&&\\
& & & CC+Pantheon+SHOES sample & &&\\
$\Lambda$CDM &$72.66\pm 0.26             $ & $0.342\pm 0.019            $&-&-&-&-\\
Power-law & $71.54^{+0.11}_{-0.093}    $ & $0.1971\pm 0.0068          $ & $-1.0284\pm 0.0096         $ & $-0.0181^{+0.011}_{-0.0082}$& $-0.343\pm 0.010           $& $0.4871\pm 0.0098          $\\
&&&&&&\\
\hline
& & & \textbf{$95\%$ limits} & & &\\
& & & CC sample & &&\\
$\Lambda$CDM &$68.8^{+1.9}_{-1.8}        $ & $0.318^{+0.068}_{-0.063}   $&-&-&-&-\\
Power-law & $71.6^{+1.0}_{-1.0}        $& $0.292^{+0.040}_{-0.040}   $ & $-1.00^{+0.18}_{-0.18}     $ & $-0.00996^{+0.0020}_{-0.0020}$ & $-0.3612^{+0.0020}_{-0.0020}$ & $0.369^{+0.094}_{-0.089}   $\\
&&&&&&\\
& & & Pantheon+SHOES sample & &&\\
$\Lambda$CDM &$72.33^{+0.55}_{-0.54}     $ & $0.383^{+0.044}_{-0.044}   $&-&-&-&-\\
    			Power-law & $71.73^{+0.16}_{-0.19}     $ & $0.190^{+0.013}_{-0.013}   $& $-1.005^{+0.020}_{-0.019}  $ & $-0.0100^{+0.0022}_{-0.0021}$ & $-0.3616^{+0.0020}_{-0.0019}$ & $0.463^{+0.012}_{-0.012}   $\\
       &&&&&&\\
       & & & CC+Pantheon+SHOES sample & && \\
       $\Lambda$CDM &$72.66^{+0.50}_{-0.53}     $ & $0.342^{+0.038}_{-0.036}   $&-&-&-&-\\
Power-law &  $71.54^{+0.19}_{-0.22}     $ & $0.197^{+0.014}_{-0.014}   $& $-1.028^{+0.020}_{-0.018}  $& $-0.018^{+0.017}_{-0.018}  $ & $-0.343^{+0.019}_{-0.020}  $ & $0.487^{+0.020}_{-0.020}   $
\end{tabular}
		   \end{ruledtabular}
		\end{table*}
\end{center}

\subsection{Information Criteria and Model Selection Analysis}

This subsection will discuss the various statistical information criteria and the model selection procedures. For this purpose, we use the Akaike information criterion (AIC) \cite{AIC1}, and the Bayesian information criterion (BIC) \cite{AIC2} to compare a set of models with their observational prediction given by dataset(s).

On the basis of information theory, the AIC addresses the problem of model adequacy. It is a Kullback-Leibler information estimator with the property of asymptotic unbiasedness. The AIC estimator is given under the standard assumption of Gaussian errors, by \cite{AIC3, AIC4}
\begin{equation}
    AIC = -2 \ln{(\mathcal{L}_{max})}+2 k + \frac{2 k\, (k+1)}{N_{tot}-k-1},
\end{equation}
where $k$ is the number of free parameters in the proposed model, $\mathcal{L}_{max}$ is the maximum likelihood value of the dataset(s) considered for analysis, and $N_{tot}$ is the number of data points. For a large number of data points, the above formula reduces to $AIC\equiv -2 \mathcal{L}_{max}+2 k $, which is a modified form of AIC. Therefore, the modified AIC criteria is convenient for all the cases \cite{AIC5}.

The BIC is a Bayesian evidence estimator, given by \cite{AIC4,AIC5,AIC6},
\begin{equation}
    BIC= -2 \ln{(\mathcal{L}_{max})}+ k \log(N_{tot}).
\end{equation}

For a given set of comparable models, we aim to rank them according to their fitting qualities with respect to the observational dataset. We use the previously studied method, in particular, the relative difference between the IC value of the given models,
\begin{equation}
\Delta IC_{model}= IC_{model}-IC_{min},
\end{equation}
where $IC_{min}$ is the minimum value of IC of the set of competing models. The $\Delta IC$ value measures the compatibility and tension between the models. According to Jeffrey's scale \cite{AIC7}, the condition $\Delta IC \leq 2$ confirms the statistical compatibility of the two models, and the model most favored by the data. The condition $2 < \Delta IC < 6$ indicates a mild tension between the two models, while the condition $\Delta IC \geq 10$ suggests a strong tension. The outputs of these tests are presented in Table~\ref{table3}.

\begin{center}
\begin{table*}[!htbp]
\caption{The corresponding $\chi^2_{min}$ of the models for each sample and the information criteria AIC, BIC for the examined cosmological models, along with the corresponding differences $\Delta IC_{model}= IC_{model}-IC_{min}$. }
		    \label{table3}
		\begin{ruledtabular}
		    \centering
		    \begin{tabular}{|c c c c c c c| }
				Model & $\chi^2_{min}$ &red. $\chi^2$ & AIC & $\Delta$ AIC & BIC & $\Delta$ BIC\\
    	\hline
     & & & CC & & &\\
    $\Lambda$CDM & 16.07 & 0.64 & 20.07 & 0 & 22.93 & 0\\
    Power-law & 16.06 & 0.64 & 28.06 & 7.98 & 36.66& 13.72\\
    &&&&&&\\
     & & & Pantheon+SHOES & & &\\
    $\Lambda$CDM & 1696.84& 1.0 & 1700.84 & 0 & 1719.15 & 0\\
    Power-law & 1683.20 & 0.99 & 1695.20 &5.63 &1727.83 & 8.6\\
    &&&&&&\\
     & & & CC+Pantheon+SHOES & & &\\
   $\Lambda$CDM  & 1712.9 & 1.0 & 1716.90 & 0 & 1735.28 & 0\\
    Power-law & 1699.33 & 0.99 & 1711.33 & 5.5 & 1744.07 &8.79\\

\end{tabular}
		   \end{ruledtabular}
		\end{table*}
\end{center}

\subsection{Numerical results}

In Tables~\ref{table2} and \ref{table3}, we have presented the numerical limits of the parameters $H_0$,\,\,$\Omega_{m0}$, $\omega_0,\,\,\, \omega_a$, $n$, and of some cosmological parameters with the 68\% and 95\% confidence levels. The constraint values on the present Hubble parameter are $71.59\pm 0.54,\, 71.733^{+0.085}_{-0.068},\,  71.54^{+0.11}_{-0.093} $
 with 68\% CL for CC, Pantheon+SHOES, CC+Pantheon+SHOES sample respectively.

 These results are consistent with recent studies (one can see the detailed discussion on $H_0$ in the reference herein \cite{H0}). Furthermore, the parameters $\omega_0,\,\,\, \omega_a$ play an important role in identifying the nature of the CPL equation of state parameter/dark energy equation of state (EoS). This EoS reduces to $\omega_0$ at $z=0$, and the constraint values on it are $-1.005^{+0.090}_{-0.090}  $, $-1.005^{+0.010}_{-0.010} $, $-1.0284^{+0.0096}_{-0.0096} $ for the respective date samples. These values are very close to the $\Lambda$CDM model.

  On the other hand $\omega_{CPL}(z)$ shows the phantom type behaviour with the constraint values on $\omega_0,\,\,\, \omega_a$ for all datasets, i.e., $\omega_{CPL}(z)<1$ always. From all these outputs, one can see that our findings confirm the existence of the present accelerated expansion of the Universe. In addition to this, we have presented the $\chi^2_{mim}$, the reduced $\chi^2_{mim}$, the AIC, BIC, $\Delta$AIC and $\Delta$BIC values in Table~\ref{table3}. From these results, we can estimate that the power law $f(Q)$ type model is a good fit to the observational datasets, as compared with the $\Lambda$CDM model. However, it shows a mild tension between models as per the information criteria analysis. Our model shows mild tension compared to $\Lambda$CDM because the modified gravity model has more degrees of freedom in the parameter spaces than $\Lambda$CDM. And, the IC values depend on the number of model parameters. These tensions may allow us to open a new path to solving the $H_0$ tension in the near future. Further, It is well-known that these types of studies in modified gravity are giving us extra degrees of freedom, which could allow us to deal with the Hubble tension precisely in the near future, and before that we have to deal with many discrepancies for example, different statistical significance, ideal number free parameters in a model. From our analysis, we can see that the $H_0$ values are a little less than $\Lambda$CDM in the case of Pantheon and CC+ pantheon samples, whereas in the case of CC, it is the opposite. As per the literature review, we have seen that $H_0$ tension is large between CMB and SNIa data analysis (for example) \cite{H0, H01}. But in our case, we can see that the $H_0$ value decreases in the case of SNIa and increases in the case of CC compared to $\Lambda$CDM. These results suggest that our model is able to reduce to $H_0$ tensions between observational samples. Moreover, we need to explore our model with other datasets to have a complete view on $H_0$ tension and its solution. In particular, we could expect that we will get a higher value $H_0$ for our model compared to $\Lambda$CDM for CMB data as per our previous data analysis. Also, the combined data analysis with observational samples may help us to reduce the $H_0$ tensions. In the near future, we hope to explore these studies. To explore more about our model, we discuss some cosmological applications in the following Section.

\section{Cosmological applications}\label{sec:5}

In this Section, we shall discuss some cosmological applications of our theoretical $f(Q)$ model, and we examine its current dynamical status. In this respect, we investigate the basic Cosmographic Parameters, the matter distribution profiles, and the dark energy types profiles, respectively.

\subsection{Cosmographic parameters}

The Cosmographic parameters are simply a Mathematical tool that considers the cosmic scale factor, and its derivatives. Using these parameters' behavior, one can investigate the present, low redshift behavior, and predict the future of the cosmological models. Therefore, we consider the profiles of the Hubble, deceleration, jerk and snap parameters to present the dynamic status of our model. Furthermore, we can write down the mathematical expressions for those parameters as follows;
\begin{eqnarray}
    q(z)=\Omega_r+\frac{1}{2}\Omega_m(z)+\frac{1+3\omega_{de}}{2} \Omega_{de}(z),
\end{eqnarray}
\begin{equation}
   j(z)= q(z) (2 q(z) + 1) + (1 + z)q'(z),
\end{equation}
\begin{equation}
   s(z)= -(1 + z) j'(z) - 2 j(z) - 3 j(z) q(z).
\end{equation}
Here, ($'$) represents one time derivative with respect to $z$.

\subsubsection{The Hubble parameter}

\begin{figure}
    \centering
    \includegraphics[scale=0.85]{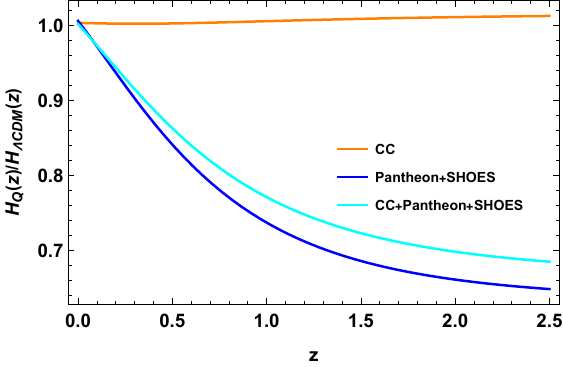}
    \caption{Evolution of the ratio $H_Q(z)/H_{\Lambda CDM}(z)$ as a function of the redshift variable $z$ for the constraint values of $H_0, \Omega_{m0},\,\omega_0, \omega_a, n,\, \gamma$ for the CC, Pantheon+SHOES, and the CC+Pantheon+SHOES samples.}
    \label{fig6a}
\end{figure}

In the previous Section, we have presented the evolution profile of the Hubble parameter with the constraint values of the free parameters. Here, we consider the ratio of $H_Q(z)/H_{\Lambda CDM}(z)$ in order to check the difference between both  models. In Fig. \ref{fig6a} we plot the redshift dependence of this ratio. For low redshifts, like, for example, for $z=0.2$, the difference between the two models is of the order of $0.0003$\%, $7.06$\%, and $5.58$\%, respectively, for the CC, Pantheon+SHOES, and CC+Pantheon+SHOES samples.

The differences between the models increase for high redshift, so that for $z=2.0$, the differences are of the order of  $0.003$\%, $27.21$\%, and $22.98$\%, respectively,  for the CC, Pantheon+SHOES, and CC+Pantheon+SHOES samples, respectively.

\subsubsection{The deceleration, jerk and snap parameters}

\begin{figure}
    \centering
    \includegraphics[scale=0.85]{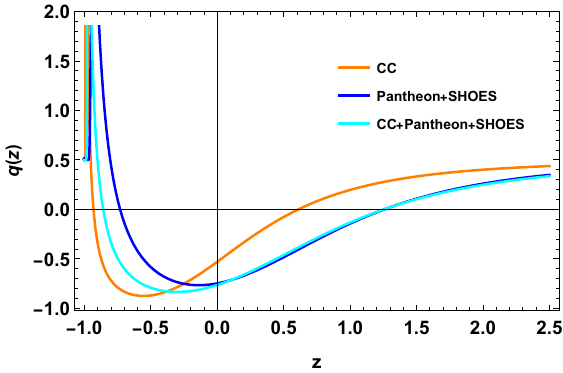}
    \caption{Evolution of the deceleration parameter as functions of the redshift variable $z$ for the constraint values of $H_0, \Omega_{m0},\,\omega_0, \omega_a, n\, \gamma$ for CC, Pantheon+SHOES, CC+Pantheon+SHOES samples.}
    \label{fig6}
\end{figure}

\begin{figure}
    \centering
    \includegraphics[scale=0.57]{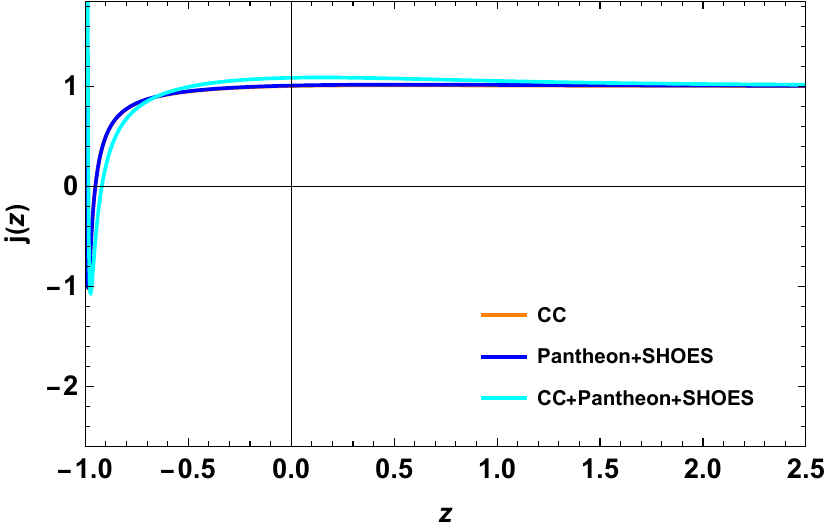}
    \caption{Evolution of jerk parameter $j$ as a function of the redshift variable $z$ for the constraint values of $H_0, \Omega_{m0},\,\omega_0, \omega_a, n\, \gamma$ for the CC, Pantheon+SHOES, and CC+Pantheon+SHOES samples.(Here the profile of the jerk parameter for CC and Pantheon+SHOES samples overlaps each other.)}
    \label{fig7}
\end{figure}

\begin{figure}
    \centering
    \includegraphics[scale=0.55]{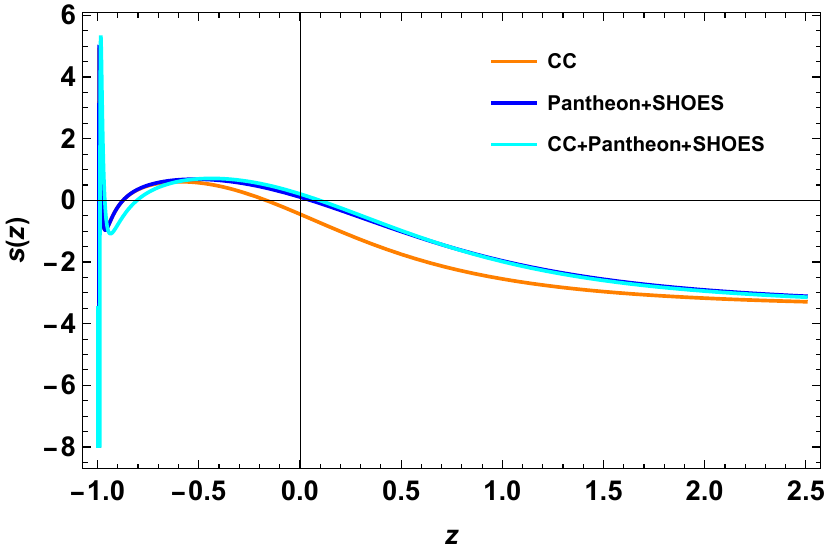}
    \caption{Evolution of the snap parameter $s$  as a function of the redshift variable $z$ for the constraint values of $H_0, \Omega_{m0},\,\omega_0, \omega_a, n\, \gamma$ for the CC, Pantheon+SHOES, and CC+Pantheon+SHOES samples.}
    \label{fig8}
\end{figure}

Furthermore, we have depicted the profiles of the deceleration, jerk, and snap parameters with the constraint values of the free parameters for the various observational datasets in Figs.~\ref{fig6}, \ref{fig7}, and \ref{fig8}, respectively.

\paragraph{The deceleration parameter.} From the redshift profile of the deceleration parameter one can see clearly that our model's evolution started from the decelerated phase, and it is currently  in an accelerating stage, after going through the matter-dominated era. In addition, we have found that the present values of the deceleration parameter $q_0= -0.532,\,\, -0.717,\,\, -0.744$ for CC, Pantheon+SHOES, CC+Pantheon+SHOES, respectively, are aligned with the recent observational results \cite{q0,q01,q02}.

\begin{figure}
    \centering
    \includegraphics[scale=0.53]{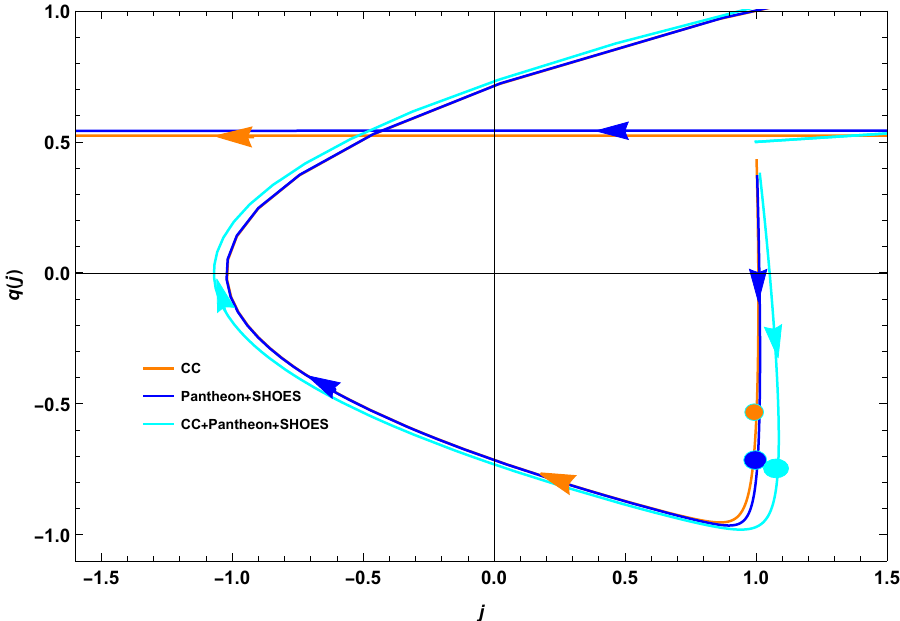}
    \caption{Parametric plot of $q=q(j)$ in the redshift range $z\in [-1,2.5]$ with the constraint values of $H_0, \Omega_{m0},\,\omega_0, \omega_a, n\, \gamma$ for the CC, Pantheon+SHOES, and the CC+Pantheon+SHOES samples. The orange, blue, and cyan color points represents the present value of the pair $(j_0,q_0)$ for respective samples.}
    \label{fig8a}
\end{figure}

\paragraph{Jerk and snap parameters.} The evolution of the jerk and snap parameters are presented for the present model in Figs.~\ref{fig7} and \ref{fig8}, respectively.  We have also obtained the parametric plot $q-j$ for the redshift range $z\in [-1,2.5]$ in Fig.~\ref{fig8a}.
In addition, we have presented $1-\sigma$ CL values of the deceleration, jerk, and snap parameters in Table~\ref{table4}. The present-day value of the jerk parameter for all the observational samples is close to the $\Lambda$CDM value.


\subsubsection{Dimensionless density parameters}

\begin{figure}[h]
    \centering
    \includegraphics[scale=0.55]{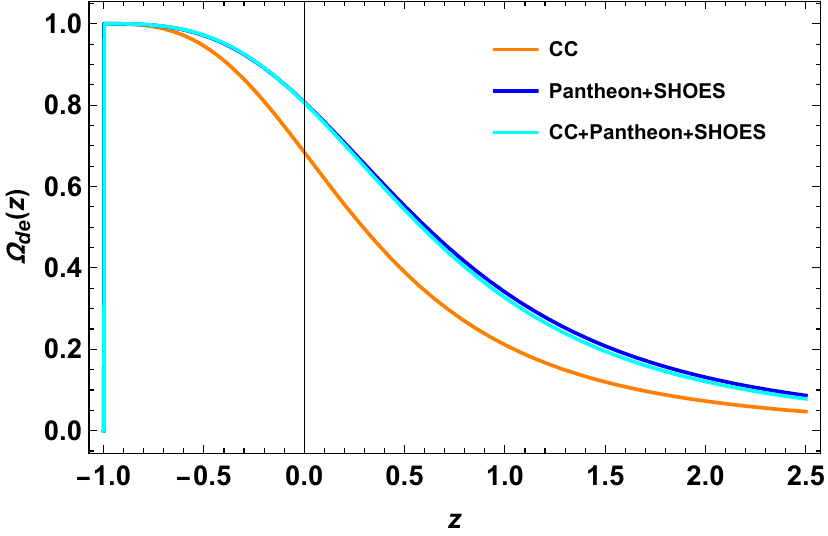}
    \caption{Profiles of the parameter of the dark energy density $\Omega_{de}$ as functions the redshift variable $z$ for the constraint values of $H_0, \Omega_{m0},\,\omega_0, \omega_a, n, \gamma$ for the CC, Pantheon+SHOES, and CC+Pantheon+SHOES samples.}
    \label{fig9}
\end{figure}

\begin{figure}[h]
    \centering
    \includegraphics[scale=0.55]{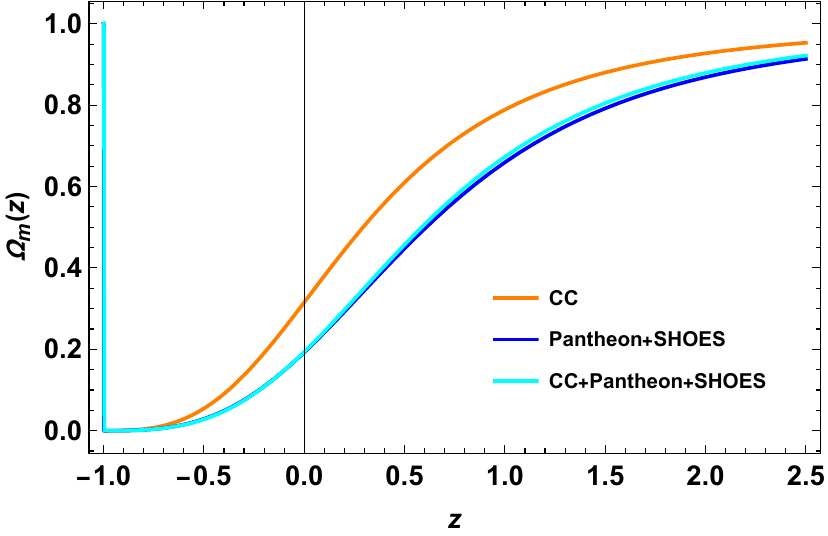}
    \caption{Profiles of the matter-energy density parameter $\Omega_{m}$ as a function of the redshift variable $z$ for the constraint values of $H_0, \Omega_{m0},\,\omega_0, \omega_a, n, \gamma$ for the CC, Pantheon+SHOES, and CC+Pantheon+SHOES samples.}
    \label{fig10}
\end{figure}

The energy density sources of our universe evolve in time,  and play a major role in characterizing its past, present, and future. Here, we have presented the evolution profiles of the dark energy density and of the matter density in Figs.~\ref{fig9} and \ref{fig10}, respectively. From those Figures, one can observe that the matter energy dominated our Universe in the early time, whereas the dark-energy density dominates in the current phase. Dark energy is also responsible for the present acceleration of the Universe. The present-day values of the dark energy density are $0.685^{+0.010}_{-0.013}$, $0.8076^{+0.0037}_{-0.0036} $, and $0.8064^{+0.0024}_{-0.0023} $ with $1-\sigma$ error for the CC, Pantheon+SHOES, and CC+Pantheon+SHOES, respectively. We also present the constraint values of the matter density and of the dark energy density in Tables~\ref{table2} and \ref{table4}, for the 68\% and 95\% confidence levels. In addition, the energy densities satisfy the relation $\Omega_m+\Omega_{de}\simeq 1$ for the entire period of their evolution. The dynamical profiles of the two fluids also suggests that dark energy will continue to dominate our Universe in the near future.

\begin{center}
\begin{table*}[!htbp]
\caption{Present-day values of the cosmological parameters $q_0$,\,\,$j_0$,\, $s_0$ and $\Omega_{de0}$ as predicted by the power law $f(Q)$ model for different data samples with 68\% confidence level. }
		    \label{table4}
		\begin{ruledtabular}
		    \centering
		    \begin{tabular}{|c c c c c|}
				Model & $q_0$& $j_0$ &$s_0$& $\Omega_{de0}$ \\
    	\hline
     &&CC sample&&\\
$\Lambda$CDM & $-0.523\pm 0.0345$& $1\pm (<\mathcal{O}(10^{-16}))$& $-0.431\pm 0.1035$& $0.682\pm 0.034$\\
    Power-law & $-0.532^{+0.077}_{-0.070}$ & $1.001^{+0.298}_{-0.258}$& $-0.439^{+0.469}_{-0.278}$ & $0.685^{+0.010}_{-0.013}$\\
    &&&&\\
      &&Pantheon+SHOES sample&&\\
    $\Lambda$CDM & $-0.4255\pm 0.033$& $1\pm (<\mathcal{O}(10^{-16}))$& $-0.7235\pm 0.099$& $0.617\pm 0.022$\\
Power-law & $-0.717^{+0.017}_{-0.017}$& $1.006^{+0.035}_{-0.035}$ & $0.108^{+0.075}_{-0.071}$& $0.8076^{+0.0037}_{-0.0036}$\\
  &&&&\\
&&CC+Pantheon+SHOES sample&&\\
$\Lambda$CDM & $-0.487\pm 0.0285$& $1\pm (<\mathcal{O}(10^{-16}))$& $-0.539\pm 0.0855$& $0.658\pm 0.019$\\
Power-law & $-0.744^{+0.015}_{-0.015}$& $1.06^{+0.023}_{-0.038}$ & $0.198^{+0.011}_{-0.413}$ & $0.8064^{+0.0024}_{-0.0023}$\\

\end{tabular}
		   \end{ruledtabular}
		\end{table*}
\end{center}

\subsubsection{$Om$ Diagnostics}

The $Om$ diagnostic is used to analyze the difference between standard $\Lambda CDM$ and other dark energy models. $Om$ is more convenient than the state-finder diagnosis \cite{Om} as it uses only the first-order temporal derivative of the cosmic scale factor. This is because it only involves the Hubble parameter, and the Hubble parameter depends on a single time derivative of $a(t)$. For the spatially flat Universe, it is defined as
 \begin{multline}
     Om(x)=\frac{\mathcal{H}(x)^2 -1}{(1+z)^3 -1},\,\, x=1+z,\mathcal{H}(x)= H(x)/H_0,
 \end{multline}
 where $z$ is the redshift, and $H_0$ is the present-day value of the Hubble parameter. For the dark energy model with the constant equation of state $\omega$,
 \begin{equation}\label{33}
     \mathcal{H}(x)= \Omega_{m0}x^3+(1-\Omega_{m0}) x^{\delta}, \,\, \delta=3(1+\omega).
 \end{equation}

 Now, we can rewrite $Om(x)$ as
 \begin{equation}
     Om(x)=\Omega_{m0}+(1-\Omega_{m0})\frac{x^{\delta}-1}{x^3-1}.
 \end{equation}
 For the $\Lambda$CDM model, we find
 \begin{equation}
     Om(x)=\Omega_{m0},
 \end{equation}
 whereas $Om(x)<\Omega_{m0}$ in phantom cosmology  with $\delta<0$, while $Om(x)>\Omega_{m0}$ in the quintessence models with $\delta>0$. These results show that: \textit{$Om(x)-\Omega_{m0}=0$}, if dark energy is a cosmological constant \cite{Om}.

 In another way, we can say that the $Om$ diagnostic gives us a \textit{null test} of the cosmological constant. As a consequence, $\mathcal{H}(x)^2$ provides a straight line against $x^3$ with a constant slope $\Omega_{m0}$ for $\Lambda$CDM, a result which can be verified by using equation \eqref{33}. For other dark energy models $Om(x)$ is curved, because
 \begin{equation}
     \frac{d\mathcal{H}^2(x)}{dx}={\rm constant}.
 \end{equation}

Furthermore, for $x_1 < x_2$, $Om(x_1,x_2)\equiv Om(x_1)-Om(x_2)=0$ in $\Lambda$CDM,  whereas $Om(x_1,x_2)\equiv Om(x_1)-Om(x_2)<0$ in phantom models, and $Om(x_1,x_2)\equiv Om(x_1)-Om(x_2)>0$ in quintessence cosmology. This test helps us with the interpretation of the observational measurements, and also, provides us a null test for the $\Lambda$CDM model. In addition to this, one can check that $Om(x)\rightarrow 0$ as $z\rightarrow -1$ for quintessence, $Om(x)$ diverges at $z<0$, suggesting the `big rip' future singularity for phantom cosmology, and $\Lambda$CDM approached towards the de Sitter spacetime at the late times.

\begin{figure}
    \centering
    \includegraphics[scale=0.55]{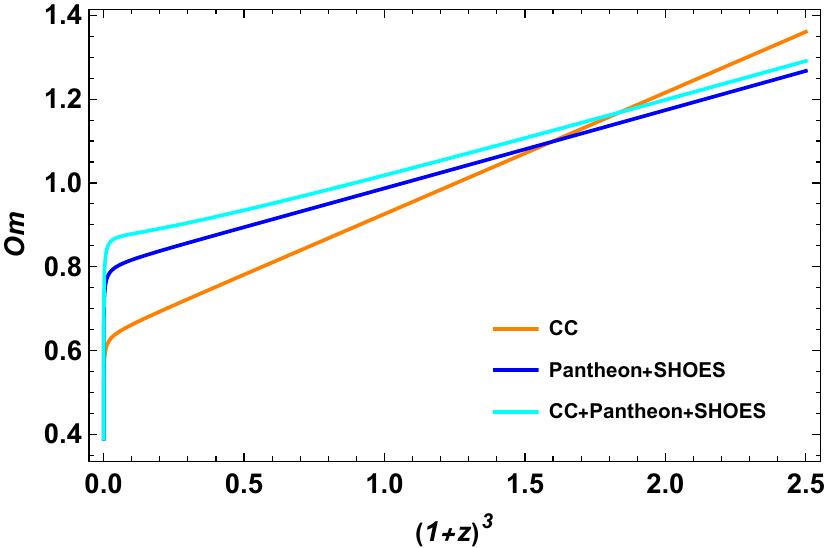}
    \caption{Profiles of the $Om$ diagnostic parameter as a function of $1+z$ for the constraint values of $H_0, \Omega_{m0},\,\omega_0, \omega_a, n,\gamma$ for the CC, Pantheon+SHOES, and  CC+Pantheon+SHOES samples.}
    \label{fig11}
\end{figure}

We have examined the $Om$ diagnostic profiles for our $f(Q)$ model with the constraint values of the parameters. We have presented our results in Fig.~\ref{fig11}. One can observe that at $z=0$, $Om(x_1,x_2)<0$, which means that the dark energy candidate of our model shows phantom-type behavior. But, in the late time, $Om(x)\rightarrow 0$ when $z\rightarrow -1$ the model has quintessence-like properties.

\section{Conclusion}\label{sec:6}

In the present paper, we have investigated in detail the cosmological properties of a particular $f(Q)$ gravity model, with the function $f(Q)$ given by $f(Q)=Q+  6\gamma H_0^2(Q/Q_0)^n$. The $f(Q)$ theory is an interesting, and fundamental approach to the description of the gravitational phenomena, in which the gravitational interaction is fully characterized by the non-metricity of the space-time $Q$, defined a general functional framework. $f(Q)$ gravity is one important component of the "geometric trinity of gravity", and offers a full and convincing alternative to the curvature description the gravitational interaction, which is used in standard general relativity, and which was so successful in the description of the gravitational interaction. From a geometric and mathematical point of view, $f(Q)$ gravity uses the Weylian extension of Riemann geometry, in which one of the fundamental prescription of this geometry, the metricity condition, is not valid anymore. The breaking of the metricity condition is thus the source of the gravitational phenomena, with the non-metricity scalar $Q$ playing an analogous role to the one played by the Ricci scalar in general relativity. In an action formulation,  for $f(Q)=Q$, we exactly recover standard general relativity. In our study we have restricted our analysis to a specific form of the function $f(Q)$, in which the deviations from standard general relativity are described by a power-law function in the non-metricity $Q$. After writing down the field equations of the $f(Q)$ theory in a general form, we have considered a specific dark energy model, in which the effective dark energy density, and its effective pressure, which are both geometric in their origin, are related by a linear, barotropic type equation of state, with a redshift dependent parameter of the EOS, $\omega _{de}=\omega _{de}(z)$. For $\omega _{de}$ we have adopted the first order CPL parameterizations, which can be extensively used for the observational testing of cosmological models. Moreover, we have restricted our basic model by imposing the energy conservation of each of the considered components of the Universe, radiation, matter, and dark energy, respectively. This procedure allows the determination of the expression of the Hubble function in terms of the three $f(Q)$ model parameters $H_0$, $\gamma$, and $n$, respectively. However, for a full comparison with the observational data, one must extend the parameter space by including the two parameters of the CPL equation of state of the dark energy.

To confront the power-law $f(Q)$ model with observations, several datasets containing cosmological data have been used. In particular, we have analyzed the model with respect to the Cosmic Chronometer (CC) dataset, as well as with the Pantheon+SHOES database. As a firs step in our investigation we have performed an MCMC analysis of the model, and obtained the optimal values of the model parameters. Then, by using these values, we have considered the general cosmological properties of this particular $f(Q)$ type theory. Generally, the MCMC analysis of all three combinations of data sets indicate a value of $n$ which is of the order of $n\approx -0.36$, or, approximately, $n=-1/3$. Hence, the dependence of the function $F(Q)$ on $Q$ is of the form $F(Q)\propto Q^{-1/3}$, that is, $F$ decreases with the increase of the nonmetricity. This interesting result may raise the problem of the explanation of this particular value of $n=-1/3$, obtained phenomenologically in the present work, by a more detailed theoretical approach. 

The deviations from standard general relativity are described by the parameter $\gamma$, which turn out to be important, with $\gamma$ having values of the order $\gamma \approx 0.45$. This indicate a large departure from the Riemannian geometry based general relativity (in the absence of a cosmological constant), but clearly indicates the possibility of the description of the dark energy in this $f(Q)$ type model. The comparison with the observational data on the Hubble parameter indicates a very good concordance between the $f(Q)$ model, $\Lambda$CDM and observations up to a redshift of $z\approx 1$, with some deviations appearing at higher redshifts. The AIC analysis also confirms the existence of a mild tension between the present model and the $\Lambda$CDM predictions, but to obtain a definite answer to this question more observational data spreading on a larger redshift range are necessary. The values of two free parameters $\omega _0$ and $\omega _a$ of the CPL type equation of state parameter of the dark energy indicate that $\omega _0\approx -1$, and hence at least at small redshifts the present model mimics a cosmological constant. The correction term $\omega _a$, giving the higher order redshift corrections is very small, of the order of $\omega _a\approx -0.01$, indicating that an effective cosmological constant, obtained from the Weyl geometric structure of the theory, gives the best description of the observational data. 

We have also performed a detailed investigation of several other cosmological parameters by using the optimal values of the $f(Q)$ model parameters. Our analysis indicate the presence of several important differences with respect to the $\Lambda$CDM model, differences whose relevance may be addressed once the precision and the number of observational data will significantly increase. For a comparison with $f(T)$ power-law model, one can see the reference \cite{ft}. The authors examined three efficient $f(T)$ models with the recent observational data in their study. The most well-fitting gravity model is the power law $f(T)$ model, which favors a minor but non-zero deviation from $\Lambda\text{CDM}$ cosmology. A Bayesian framework is used to study $f(T)$ gravity, considering both background and perturbation behavior simultaneously \cite{ft1}. The authors analyzed three viable $f(T)$ gravity models and showed that those $f(T)$ models can appropriately describe the $f\sigma8$ data. In the above studies in $f(T)$ gravity, authors have tested various $f(T)$ models againt the observational data and then compared with the $\Lambda$CDM. Whereas in our study, we have not only confronted our model against the observational datasets but also used the outputs to explore the various cosmological applications starting from the cosmographics parameters, energy densities to the dark energy profile of our model. Further, we have explored the dark energy equation of state ($\omega_{de}$) precisely comparing with $\Lambda$CDM model.

The $f(Q)$ theory of gravity can also be extended to include, together with the ordinary matter, scalar or other physical fields in the action. The present power-law $f(Q)$ model may have some other possible applications, like, for example, to consider inflation in the presence of both scalar fields and nonmetricity, an approach that may lead to the formulation  of a new view on the gravitational, geometrical and cosmological processes that did shape and influence the dynamics of the very early Universe. Another major topic of research would be the investigation of structure formation in the power-law $f(Q)$ theory which could be done with the use of a background cosmological metric, obtained by solving exactly or approximately the cosmological evolution equations. In this case the BAO, SNIa, and CMB shift parameter data could be investigated to obtain important physical and cosmological constraints for the power law $f(Q)$ model. This approach may lead to a detailed investigation and analysis of the cosmic structure formation processes, by providing  a new perspective on these processes, and on the role of Weyl non-metricity. Another direction of research would be to obtain the Newtonian and the post-Newtonian approximations  of the present power-law $f(Q)$ gravity, and to find out what constraints the local classic Solar System tests impose on the free parameters of the theory, and if these constraints are consistent with the cosmological observations. The Newtonian and the post-Newtonian limits may also prove to be extremely useful in obtaining physical constraints from a large body of  astrophysical observations.

To conclude, in our work we have developed a particular version of the $f(Q)$ theory, with the functional form of $f$ given by a simple power law function,  and we have proven its consistency with the cosmological observations, and as an important theoretical tool for the understanding of the accelerating expansion of the Universe. The obtained results also suggests the necessity of the study of further extensions and generalizations of this simple  $f(Q)$  type model. Our results have shown that the present poser-law model  may represent an interesting geometric alternatives to dark energy, going below the Riemannian mathematical structure of general relativity, and in which the non-metric properties of the space-time may offer the clue for a deeper understanding of the gravitational interaction. In the present study we have proposed  some basic theoretical tools, and observational/statistical procedures for the investigation of the basic geometric aspects of gravity, from a different perspective than the Riemannian one,  and of their cosmological applications.

\section*{Acknowledgements} We are very grateful to the anonymous referee for comments and suggestions that have significantly improved our work in terms of research quality, and presentation. S.M. acknowledges Transilvania University of Brasov for Transilvania Fellowship for postdoctoral research. SP \& PKS  acknowledges the National Board for Higher Mathematics (NBHM) under the Department of Atomic Energy (DAE), Govt. of India for financial support to carry out the Research project No.: 02011/3/2022 NBHM(R.P.)/R \& D II/2152 Dt.14.02.2022. The work of TH is supported by a grant of the Romanian Ministry of Education and Research, CNCS-UEFISCDI, project number PN-III-P4-ID-PCE-2020-2255 (PNCDI III).

\end{document}